\documentclass[preprint,12pt]{elsarticle}
\usepackage[left=2.8cm, right=2.8cm, top=3cm]{geometry}
\usepackage{graphicx}
\usepackage{amssymb}
\usepackage{amssymb}% http://ctan.org/pkg/amssymb
\usepackage{pifont}% http://ctan.org/pkg/pifont
\usepackage{graphicx}
\usepackage{lineno}
\usepackage{array}
\usepackage{amsmath}
\usepackage{mathtools}
\DeclarePairedDelimiter\abs{\lvert}{\rvert}%
\DeclarePairedDelimiter\norm{\lVert}{\rVert}%
\usepackage{comment}
\usepackage{float}
\usepackage{setspace}
\usepackage{array}
\usepackage{booktabs}

\makeatletter
\let\oldabs\abs
\def\abs{\@ifstar{\oldabs}{\oldabs*}}
\let\oldnorm\norm
\def\norm{\@ifstar{\oldnorm}{\oldnorm*}}
\makeatother

\journal{Journal}

\begin{document}
\doublespacing
\begin{frontmatter}

\title{Assessment of WRF model parameter sensitivity for high-intensity precipitation events during the Indian summer monsoon}
\let\today\relax

\author{Sandeep Chinta}
\author{Yaswanth Sai J}
\author{C Balaji\corref{cor1}}
\ead{balaji@iitm.ac.in}
\cortext[cor1]{Corresponding author}
\address{Indian Institute of Technology Madras}

\begin{abstract}
%% Text of abstract
Default values for many model parameters in Numerical Weather Prediction models aretypically adopted based on theoretical or experimental investigations by scheme designers. Short-range forecasts are substantially affected by the specification of parameters in the Weather Research and Forecasting (WRF) model. The presence of a multitude of parameters and several output variables in the WRF model renders appropriate parameter value identification quite challenging. The objective of the current study is to reduce the uncertainty in the model outcomes through the recognition of parameters which most strongly influence the model performance using a Global Sensitivity Analysis (GSA) method. Morris one-step-at-a-time (MOAT), a GSA method, is used to identify the sensitivities of 23 chosen tunable parameters corresponding to seven physical parameterization schemes of the WRF model. The sensitivity measures (MOAT
mean and standard deviation) are evaluated for eleven output variables, which are simulated by the WRF model, with respect to different parameters. Twelve high-intensity four-day precipitation events during the Indian summer monsoon (ISM) for the years 2015, 2016, and 2017 over the monsoon core region in India are considered for the study. Though the sensitivities of each parameter vary from one model output variable to the other, overall results suggest a general trend. Nine out of 23 parameters have a considerable effect on the outcome of the simulations, whereas the influence of a few parameters is noticeably small on the WRF model performance.\\
\end{abstract}

\begin{keyword}
global sensitivity analysis; Morris one-at-a-time; parameter sensitivity; WRF model; Indian summer monsoon
\end{keyword}

\end{frontmatter}

%\linenumbers

%% main text
\section{Introduction}\label{S:1}
Indian summer monsoon (ISM) or the southwest monsoon is one of the oldest global monsoon phenomena. ISM is irregular and erratic, whose vagaries remarkably affect India both agriculturally and economically \citet{krishna2004climate}. Bursting refers to the quirk increase in mean daily rainfall and is a characteristic of ISM. Quintessentially, ISM bursts at the beginning of June and slowly withdraws towards the start of October. Therefore, the ISM rainfall (ISMR) is more often than not referred to as June-September rainfall. ISMR contributes to more than 80 \% of the annual rainfall in India \cite{rajeevan2013study}. The monsoon core region (MCR; 69$^{\circ}$E to 88$^{\circ}$E and 18$^{\circ}$N to 28$^{\circ}$N) is a critical zone where ISMR plays a crucial role \cite{rajeevan2010active} corresponding to mean monsoon and intraseasonal variability. Besides being a conspicuous contributor to the food security and water resources of South Asian region, the ISM rainfall, representing an abundant heat source, has a significant impact on the global climate and general circulation \cite{rao1976southwest,chen2018multiscale}. Consequently, accurate predictions of the ISMR over the monsoon core region are critical not only for the water resource management in India but also for a superior seasonal forecast across the globe. \\

Mesoscale Numerical Weather Prediction (NWP) models are imperative in providing accurate seasonal and long-range predictions of the summer monsoon. Over the years, NWP models are significantly improved with refined temporal and spatial resolution, more precise dynamical methods to represent climate models, ability to simulate large-scale systems, and longer lead times in forecasting \cite{tripoli1982colorado,dudhia1993nonhydrostatic,janjic1994step,lau2009simulation,chen2011modeling,grell1994description}. The Advent of Advanced Weather Research and Forecasting (WRF) model, because of its ability to assimilate modules developed separately \cite{dudhia2014history}, is a breakthrough in mesoscale modelling. The uncertainties exhibited by NWP systems such as a WRF model can be due to four primary sources of errors: (i) the determination of initial and lateral boundary conditions,  (ii) the representation of physical processes, (iii) the specification of model input variables, and (iv) the computational precision used in the model. All the causes of uncertainties are required to be quantified and reduced for better model performance. \\

Generally, the initial and lateral boundary conditions for a WRF model are provided by large scale global analyses but with the constraints of low resolution and insufficient depiction of regional mesoscale characteristics. Different data assimilation techniques, which ingest the localized observational data, such as variational and ensemble data assimilation methods are integral to WRF and have been shown to have a considerable effect on the rainfall forecast in India \cite{routray2010simulation,dhanya2014improved}.\\

Accurate depiction of all the physical process occurring in the atmosphere is impractical to incorporate in a numerical model even with the state-of-the-art super-computing technology. Therefore, NWP developers use parameterization schemes. WRF provides a plethora of advanced physics parameterization schemes \cite{skamarock2005description} to represent the physical processes (related to the surface layer, cumulus, microphysics, short-wave and long-wave radiation, ocean model, land surface and planetary boundary layer). Any reasonable combination of physics schemes compounds to a different version of the WRF model. In literature, many studies have explored a different set of schemes for various physical processes like cyclones, severe storms, and semi-arid precipitation \cite{sandeep2018impact,soni2014performance,mohan2011analysis,halder2015effect} over different Indian regions. A few studies have also investigated the best possible physics schemes, with significant improvement in the forecast performance, for the Indian summer monsoon \cite{srinivas2013simulation,anil2010evaluation}. \\

Nonetheless, to obtain reliable seasonal predictions that are close to the truth, refined initial and boundary conditions and the employment of appropriate parameterization schemes alone are not adequate. The appropriate specification of model input variables within the model physics considered plays a pivotal role in the reliability of WRF model simulations. Each parameterization scheme has a vast number of parameters whose default values are fixed based on experimental or abstract investigations by the scheme designers. Some researchers have investigated the different possible parameter sets for various regions using trial and error based techniques \cite{allen1999yourself,knutti2002constraints} but the effectiveness of this approach is contingent on the skill of the model observer. Many examined the utility of inverse techniques, a more objective strategy, through parameter adjustment by pruning the objective function to match the simulation results with those of observations. The presence of a vast number of tunable parameters,  numerous atmospheric output variables from the simulations and computationally expensive NWP models make suitable parameter specification using conventional inverse routines extremely demanding. \\

A systematic and mathematically robust approach for parameter uncertainty quantification is required, which identifies the model input variables that have a considerable impact on the model outcome and later enables the adjustment of parameters to obtain the ideal solution. Sensitivity analysis (SA) is the most frequently utilized statistical tool to recognize the most important parameters when there is uncertainty involved. There are various SA studies on uncertainty quantification of parameters which concentrates on parameters of a particular scheme \cite{liu2004exploring,bastidas2006parameter,hou2012sensitivity,li2013assessing}. A few studies comprehensively considered many parameters across various physics schemes \cite{di2015assessing,quan2016evaluation}. But these studies are conducted over small regions. The objective of the present study is to apply the SA methods to parameters corresponding to seven different physics schemes on a larger region (Monsoon core region) when the output variables are evaluated for sensitivities concerning high-intensity precipitation events during the Indian summer monsoon. \\

This paper is organized as follows. Section \ref{S:2} gives an introduction to the Morris one at a time SA technique. Section \ref{S:3} gives a detailed description of the parameters, physics schemes, and events considered for the SA. Section \ref{S:4} lists out the critical results and inferences related to the sensitivity measures.

\section{Sensitivity analysis method}
\label{S:2}

Sensitivity Analysis (SA) is defined as the recalculation of outcomes of the model (numerical or otherwise) under alternative assumptions, owing to the uncertainties in the model inputs, to ascertain the influence of different input variables (i.e., parameters). SA finds its application, among the other things, in robustness evaluation, uncertainty reduction, error search, and model simplification. However, the prime focus of the present study is to identify the model input variables that cause substantial uncertainty in the model outcome, which then can be further researched for improved robustness of the model. Sensitivity analysis employed in the current study comprises three significant steps: (i) selection of an accurate model to be used and the identification of the adjustable model input variables, (ii) quantification of uncertainties (i.e., ranges and probability distributions) in each parameter, (iii) choice of a suitable sensitivity analysis method to evaluate parameter sensitivity.\\

The WRF model applied to the Indian summer monsoon, and a set of 23 parameters, with uniform distribution, corresponding to various physics schemes are considered for the SA in the present study. The ranges and variations of all the model input variables are presented in the next section. The usage of computationally expensive variance-based methods is not feasible as the number of model input variables is too large. At the same time, the application of group techniques is not required as the number of parameters considered for the parametric study is not large enough. The Morris one-at-a-time (MOAT) technique, also known as Elementary Effects method \cite{morris1991factorial} is an ideal technique for the number of parameters under investigation and is selected as the sensitivity analysis method in this study. The sensitivity measures obtained from this method distinguishes the parameters based on whether the effect is (a) insignificant, (b) linear and additive (c) nonlinear or involves interactions with other model input variables \cite{saltelli2008global}. \\

Before applying the MOAT method, all the ranges of \textit{k} number of parameters have to be mapped to the parent domain [0,1]. One way of doing this is by obtaining the cumulative distribution function (CDF) for the given parameter distribution, as CDF is bounded by 0 and 1 and is also uniformly distributed over [0,1]. When the parameter distribution itself is uniform, CDF will be a linear mapping from the parameter domain to the parent domain. Then the parent domain corresponding to each parameter is divided into \textit{p} equally spaced intervals. This process creates an input space $\Omega$ with a \textit{k} dimensional unit cube discretized into \textit{p} selected levels along each dimension. An initial parameter set or a base value $x^0$ = ($x^0_1$ , $x^0_2$, $x^0_3$,...., $x^0_k$) is arbitrarily chosen in the \textit{p}-level grid $\Omega$. To generate the trajectory point $x^i$, a $\Delta_i$ value is randomly selected from the set \{$\dfrac{1}{(p-1)}$, $\dfrac{2}{(p-1)}$, ... , $\dfrac{(p-2)}{(p-1)}$\} and is either added to or subtracted from one of the components of the $x^{i-1}$ vector such that $x^i$ still lies in $\Omega$. Finding all the $k+1$ points, including the initial parameter set, completes the full MOAT trajectory. The selection of the component to which the $\Delta$ value is added should be such that all the components are perturbed at least once when the entire trajectory is completed. \\

The MOAT method requires \textit{r} random replications of MOAT paths to obtain reliable SA results. We used a sampling strategy proposed by \cite{campolongo2007effective} that facilitates a scattered distribution of MOAT trajectories in the input domain. A large number of different MOAT trajectories are initially generated; the subset of \textit{r} trajectories with the highest \textit{spread} is chosen for further analysis. The \textit{spread} of \textit{r} trajectories is the square root of the sum of squares of all the $^rC_2$ distances. The distance, $d_{ml}$, between a set of trajectories \textit{m} and \textit{l} is:

\begin{equation}
	d_{ml}=\begin{cases}
	\Sigma_{i=0}^{k}\Sigma_{j=0}^{k}\sqrt{\Sigma_{z=1}^{k} [x_{z}^{i}(m) - x_{z}^{j}(l)]^2} &  m\neq l \\
	0 & otherwise
	\end{cases}
\end{equation}

where $x_z^{(i)}(m)$ is the $z^{th}$ component of the $i^{th}$ point corresponding to the $m^{th}$ trajectory. The elementary effect or the gradient corresponding to the $i^{th}$ parameter, when $x^l$ and $x^{l+1}$ of $j^{th}$ trajectory differ by $\Delta_l$ in their $i^{th}$ component is

\begin{equation}
	EE_i^j = \dfrac{[y(x^{(l)}_1 , x^{(l)}_2,.., x^{(l)}_i+\Delta_l,.., x^{(l)}_k)-y(x^{(l+1)}_1 , x^{(l+1)}_2,.., x^{(l+1)}_i,.., x^{(l+1)}_k)]}{\Delta_l}
\end{equation}

where \textit{y(x)} is the objective function, an output variable in the context of a WRF model. The elementary effect represents the effect on the model objective function \textit{y(x)} due to the $\Delta_l$ change in the $i^{th}$ parameter. When all the \textit{r} elementary effects corresponding to the $i^{th}$ parameter are computed, the sensitivity measures $\mu_i$ and $\sigma_i$ can be calculated.

\begin{eqnarray}
	\mu_i &=& \dfrac{1}{r}\sum_{j=1}^{r}\abs{EE_i^j} \\
	\sigma_i^2 &=& \dfrac{1}{r-1}\sum_{j=1}^{r}{(EE_i^j-\mu_i)}^2
\end{eqnarray}
$\mu_i$ is the mean of the absolute values of elementary effects of parameter \textit{i} and it depicts the parameter's overall impact. A larger mean value of a parameter indicates a higher influence on the model output. $\sigma_i^2$ is the variance ($\sigma$ being the standard deviation) of elementary effects of parameter \textit{i}; larger the variance of a parameter is the more is the interaction with other parameters in influencing the outcome of the simulation.

\section{Design of experiments}
\label{S:3}
\begin{figure}
    \centering
    \includegraphics[scale=0.4]{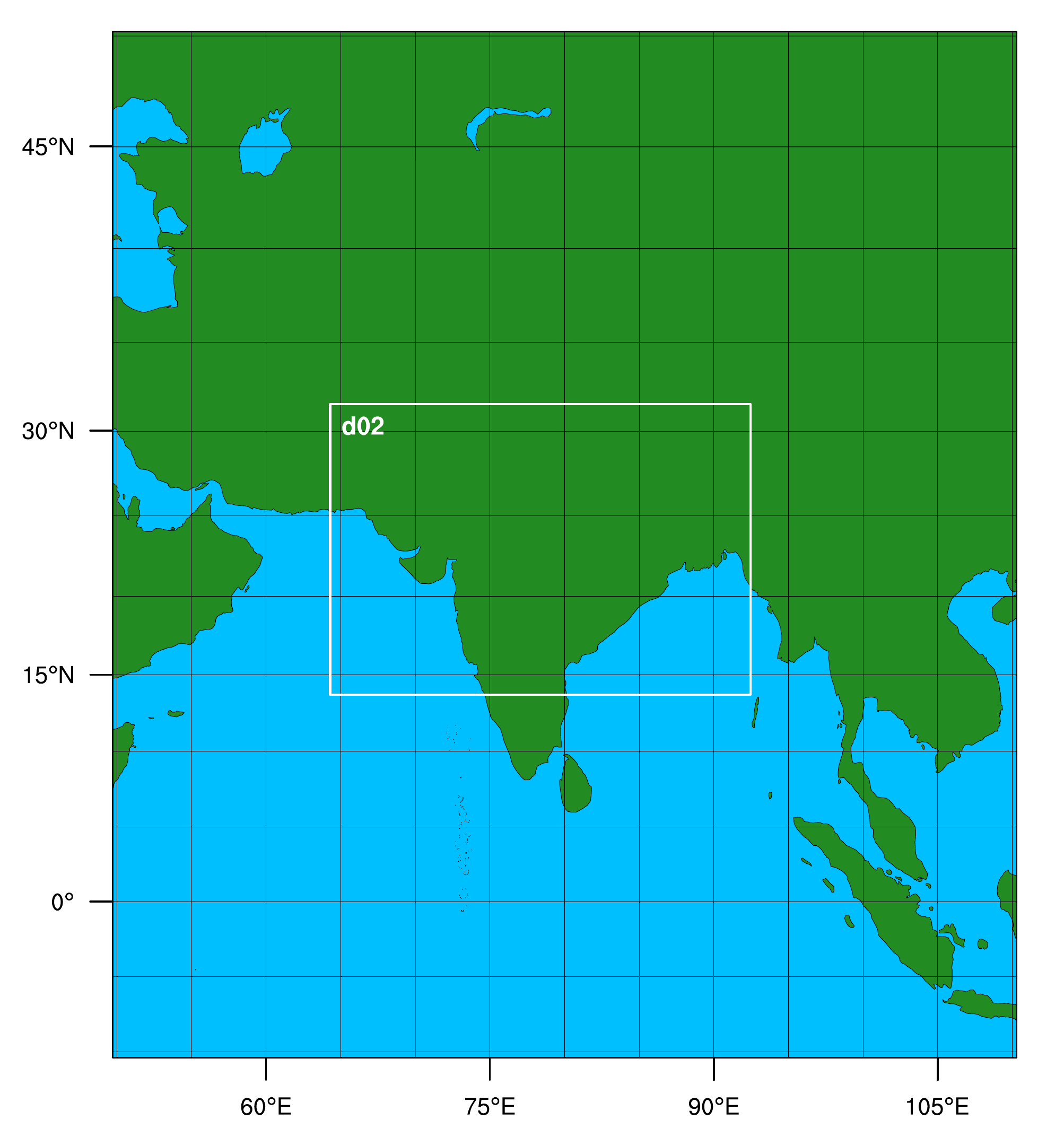}
    \footnotesize
    \caption{Configuration of model domains used in the present study}
    \label{fig:my_label}
\end{figure}

The model used in this parametric study is the Advanced Research WRF (WRF-ARW) model version 3.7.1, a non-hydrostatic compressible mesoscale numerical weather prediction system developed by the National Centre for Atmospheric Research (NCAR). The study area chosen is a two$-$grid nested model with the outer domain d01 as shown in Figure \ref{fig:my_label} \cite{ncar2017version} enclosing the Indian Subcontinent and the regions surrounding it; the inner nested domain is the Indian monsoon core region as shown in Figure \ref{fig:my_label}. The central point for d01 is $80^{o}$N, $23^{o}$E. The outer domain (d01) is composed of 188 points in the east$-$west direction and 213 points in the north$-$south direction with a spatial resolution of 36 km. The inner domain (d02) is composed of 262 points in the east$-$west direction and 181 points in the north$-$south direction with a spatial resolution of 12 km. The time$-$step for integration is 120 s and 40 s for the domains d01 and d02 respectively. With a higher resolution in the boundary layer, the vertical profile is partitioned into 40 sigma ($\sigma$) layers from the land surface to 50 hPa level in the atmosphere. The six-hourly NCEP FNL (National Center for Environmental Prediction Final) Operational Global Analysis data \cite{cisl_rda_ds083.2} at $1^{\circ}\times 1^{\circ}$ resolution provided the initial and boundary conditions for the WRF model.\\

\begin{figure}
    \centering
    \includegraphics[scale=0.85]{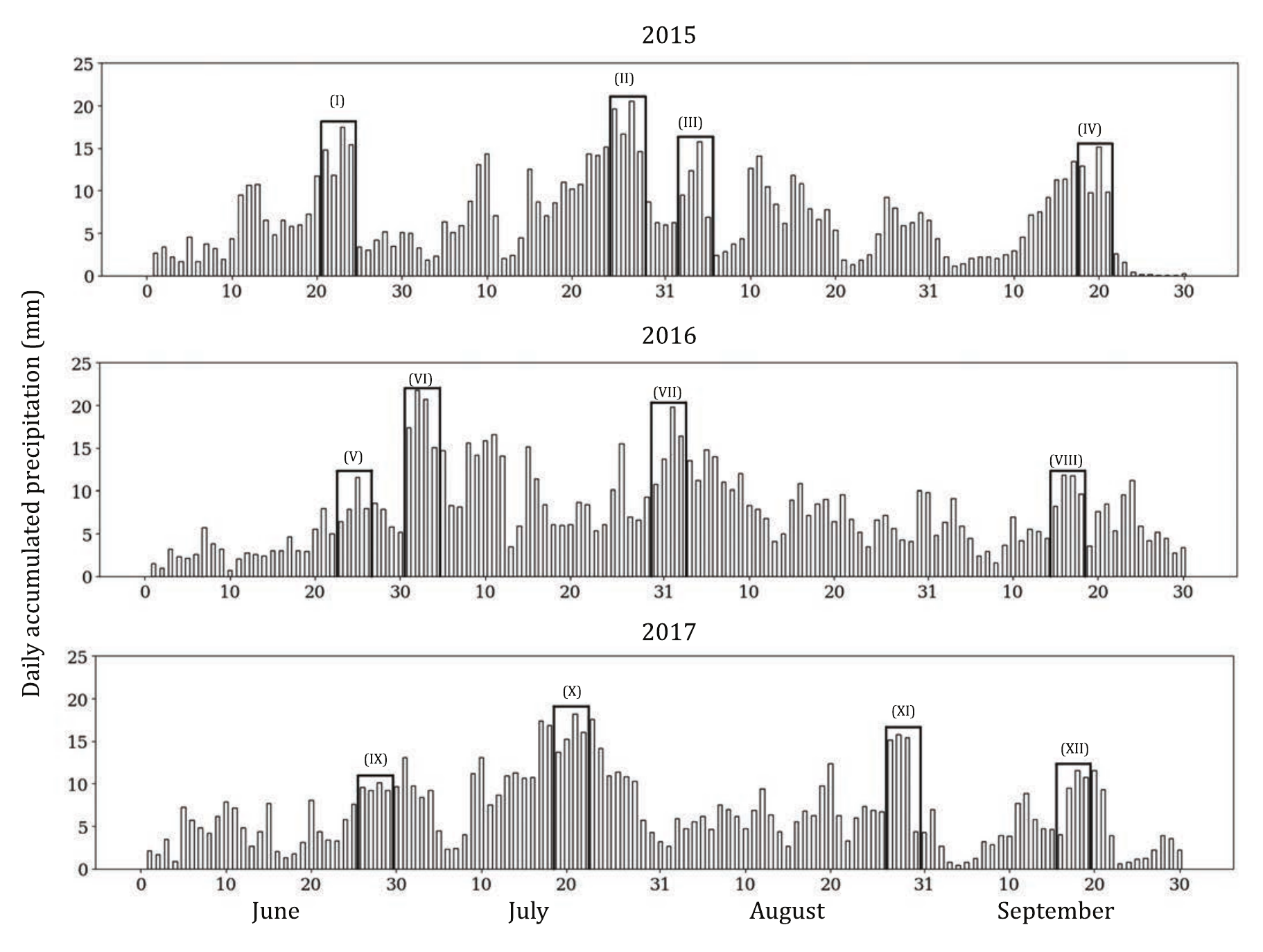}
    \caption{Daily regional average precipitation in the monsoon core region. Bold faced black boxes show the events that are simulated.}
    \label{fig:my_label2}
\end{figure}

The focus of this study is on the high-intensity rainfall instances during the Indian summer monsoon (ISM). Accordingly, 12 four day precipitation events over June, July, August, and September are chosen for the sensitivity analysis as shown in Figure \ref{fig:my_label2}. Each event contains the day with the highest rainfall in the monsoon core region in their respective month; the precipitation data is given by the Global Precipitation Measurement (GPM) Integrated Multi-satellitE Retrievals for GPM (IMERG). For events (I), (II), (III), and (IV) the simulation dates are from 21$^{\textnormal{st}}$ June to 24$^{\textnormal{th}}$ June, 25$^{\textnormal{th}}$ July to 28$^{\textnormal{th}}$ July, 2$^{\textnormal{nd}}$ August to 5$^{\textnormal{th}}$ August, and 18$^{\textnormal{th}}$ September to 21$^{\textnormal{st}}$ September in the year 2015 respectively. For events (V), (VI), (VII), and (VIII) the simulation dates are from 23$^{\textnormal{rd}}$ June to 26$^{\textnormal{th}}$ June, 1$^{\textnormal{st}}$ July to 4$^{\textnormal{th}}$ July, 30$^{\textnormal{th}}$ July to 2$^{\textnormal{nd}}$ August, and 15$^{\textnormal{th}}$ September to 18$^{\textnormal{th}}$ September in the year 2016 respectively. For events (IX), (X), (XI), and (XII) the simulation dates are from 26$^{\textnormal{th}}$ June to 29$^{\textnormal{th}}$ June, 19$^{\textnormal{th}}$ July to 22$^{\textnormal{nd}}$ July, 27$^{\textnormal{th}}$ August to 30$^{\textnormal{th}}$ August, and 16$^{\textnormal{th}}$ September to 19$^{\textnormal{th}}$ September in the year 2017 respectively. \\
The physical processes represented in WRF are near-surface physics, cumulus convection, microphysics, short-wave and long-wave radiative transfer, land-surface physics, and planetary boundary layer physics. A great variety of parameterization schemes are available for each physical process. Many adjustable parameters are present in each parameterization scheme. Most of the parameterization schemes used in this study are chosen from \cite{ratnam2017sensitivity}, where a sensitivity of physical parameterization schemes concerning the Indian summer monsoon was performed and the best set of schemes was proposed. The microphysics scheme alone was chosen differently from that set as WSM6 scheme represents more categories of hydrometeors compared to the WSM3 scheme. The schemes used in this study are shown in Table \ref{table:1}. \\
\bgroup
\def\arraystretch{1.5}
\begin{table}
\centering
\caption{Overview of parameterization schemes used in the WRF model setup}
\vspace{1mm}
\label{table:1}
\scriptsize
\begin{tabular}{m{4cm} m{5cm}}
\hline
Physical process & Specific scheme \\ 
\hline
Surface layer & MM5 Monin$-$Obukhov scheme \cite{dudhia2005psu} \\
Cumulus & Kain$-$Fritsch Eta scheme \cite{kain2004kain}\\
Microphysics & WSM 6 single-class scheme \cite{hong2006wrf}\\
Short-wave radiation & Dudhia scheme \cite{dudhia1989numerical}\\
Long-wave radiation & RRTM scheme \cite{mlawer1997radiative}\\
Land surface & Noah land surface model scheme \cite{chen2001coupling}\\
Planetary boundary layer & Yonsei University (YSU) scheme \cite{hong2006new}\\
\hline
\end{tabular}
\end{table}
\egroup

A set of 23 tunable parameters from the physics parameterization schemes in Table \ref{table:1} are used for analysis in this study, similar to \cite{quan2016evaluation}. By no means is this set exhaustive; nonetheless, the present parametric study can be considered as a precursor for further investigation of the sensitivity of other potential parameters for ISM. The overall list of parameters and their allowable ranges of variations are presented in Table \ref{table:2}.\\ 
\bgroup
\def\arraystretch{1.75}
\begin{center}
\begin{table}
\caption{Set of adjustable parameters in the WRF model}
\vspace{1mm}
\label{table:2}
\scriptsize
\begin{tabular}{l m{1.7cm} c c c m{7.1cm}} 
%l m{1.7cm} c c c m{7.1cm}
\hline
Index & Scheme & Parameter & Default & Range & Description\\ 
\hline
P1 & Surface layer & xka & 2.4e-5 & [1.2e-55e-5] & The parameter for heat/moisture exchange coefficient (sm$^{-2}$)\\
P2 && czo\_fac & 0.0185 & [0.01 0.037] & The coefficient for converting wind speed to roughness length over the water\\
\hline
P3 & Cumulus & pd & 1 & [0.5 2] & The multiplier for downdraft mass flux rate\\
P4 & & pe & 1 & [0.5 2] & The multiplier for entrainment mass flux rate\\
P5 & & ph\_usl & 150 & [50 350] & Starting height of downdraft over USL (hPa)\\
P6 & & timec & 2700 & [1800 3600] & Mean consumption time of CAPE (s)\\
P7 & & tkemax & 5 & [3 12] & The maximum turbulent kinetic energy (TKE) value in sub-cloud layer (m$^{2}$s$^{-2}$)\\
\hline
P8 & Microphysics & ice\_stokes & 14900 & [8e3 3e4] & Scaling factor applied to ice fall velocity (s$^{-1}$)\\
P9 & & n0r & 8e+6 & [5e6 1.2e7] & Intercept parameter of rain (m$^{-4}$)\\
P10 & & dimax & 5e-4 & [3e-4 8e-4] & Limited maximum value for the cloud-ice diameter (m)\\
P11 & & peaut & 0.55 & [0.35 0.85] & Collection efficiency from cloud to rain auto conversion\\
\hline
P12 & Short-wave & cssca\_fac & 1e-5 & [5e-6 2e-5] & Scattering tuning parameter (m$^{2}$kg$^{-1}$)\\
P13 & & Beta\_p & 0.4 & [0.2 0.8] & Aerosol scattering tuning parameter (m$^{2}$kg$^{-1}$)\\
\hline
P14 & Long-wave & Secang & 1.66 & [1.55 1.75] & Diffusivity angle for cloud optical depth computation\\
\hline
P15 & Land surface & hksati & 1 & [0.5 2] & Multiplier for hydraulic conductivity at saturation\\
P16 & & porsl & 1 & [0.5 2] & Multiplier for the saturated soil water content\\
P17 & & phi0 & 1 & [0.5 2] & Multiplier for minimum soil suction\\
P18 & & bsw & 1 & [0.5 2] & Multiplier for Clapp and Hornbereger `b' parameter\\
\hline
P19 & PBL* & Brcr\_sbrob & 0.3 & [0.15 0.6] & Critical Richardson number for boundary layer of water\\
P20 & & Brcr\_sb & 0.25 & [0.125 0.5] & Critical Richardson number for boundary layer of land\\
P21 & & pfac & 2 & [1 3] & Profile shape exponent for calculating the momentum diffusivity coefficient\\
P22 & & bfac & 6.8 & [3.4 13.6] & Coefficient for Prandtl number at the top of the surface layer\\
P23 & & sm & 15.9 & [12 20] & Countergradient proportional coefficient of non-local flux of momentum\\
\hline
\end{tabular}
\begin{flushright}
\scriptsize
*Planetary Boundary Layer
\end{flushright}
\end{table}
\end{center}
\egroup

Morris One at a Time (MOAT) method is used for the sensitivity analysis of all the 23 parameters. The model with 23 independent inputs varies in the 23-dimensional unit cube across a four-level grid (k=23, p=4). For the number of grid levels p=4, previous studies \cite{campolongo1997sensitivity,campolongo1999tackling,saltelli2000sensitivity} suggested a total of 10 replications. A large number ($\sim$500) of different MOAT trajectories are initially generated. The best 10 (r=10) trajectories are then selected out of the 500 trajectories by maximizing the spread among them. One complete WRF simulation requires approximately 32 CPU hours. Based on the MOAT algorithm, a total number of 240 ((23 + 1) $\times$ 10) WRF simulations are conducted for each event. Therefore considering 12 events, a total of 240 $\times$ 32 $\times$ 12 = 92160 CPU hours are utilized for this study. \\

Outside of these, the WRF model with the default parameter set is run for all the twelve precipitation events. All the eleven output variables are extracted after each successful run. Once the whole set of 240 simulations corresponding to a particular event are entirely run, elementary effects are calculated for every parameter with each of the model output variables as an objective function. With eleven model outcomes and r = 10 MOAT replications, altogether 110 gradient values are generated for every single parameter. Subsequently, both the sensitivity measures, MOAT mean and standard deviation are evaluated, as explained in the preceding section. After the completion of all the simulations, overall sensitivity is computed for all the parameters by averaging the sensitivity measures of all the 12 events. \\

The output variables from simulations have to be validated with the observed data to verify the accuracy of the simulations. In this study, IMERG daily accumulated precipitation data \cite{huffman2017gpm} at $0.1^{\circ}\times 0.1^{\circ}$ resolution is used to validate the rainfall data and ERA-Interim reanalysis data \cite{dee2011era} at $0.125^{\circ}\times 0.125^{\circ}$ resolution is used to validate surface pressure, surface air temperature and wind speed. Root Mean Square Error (RMSE) is used as the objective function to verify the closeness of the simulation to the observation.  

\begin{equation}
    RMSE = \sqrt{\dfrac{\sum_{i=1}^{N}\sum_{t=1}^{T}(sim_i^t - obs_i^t)^2}{N\times T}}
\end{equation}

\begin{flushleft}
where $sim_i^t$ and $obs_i^t$ are the simulated and observed values of a variable at grid point $i$ and time $t$ respectively. N is the total number of grid points of domain d02 in MCR and T is the number of days of simulation. 
\end{flushleft}

\section{Results and discussion}
\label{S:4}

Several meteorological variables such as daily accumulated precipitation (PR), relative humidity (RH), surface air temperature (SAT), wind speed (WS), surface air pressure (SAP), downward short-wave radiative flux (DSWRF) and downward long-wave radiative flux (DLWRF) and also atmospheric variables such as total precipitable water (TPW), planetary boundary layer height (PBLH), cloud fraction (CF) and outgoing long-wave radiation at the top of the atmosphere (OLR) from the model simulations are used to evaluate the sensitivity of the selected parameters. The mean and standard deviation values obtained form the basis of the analysis of the sensitivity of parameters. A higher value of the  MOAT-mean implies that the parameter is more sensitive, whereas a higher MOAT standard deviation value indicates that the parameter is more dependent on other parameters to affect the output. \\
\begin{figure}
\centering
    \includegraphics[scale=1]{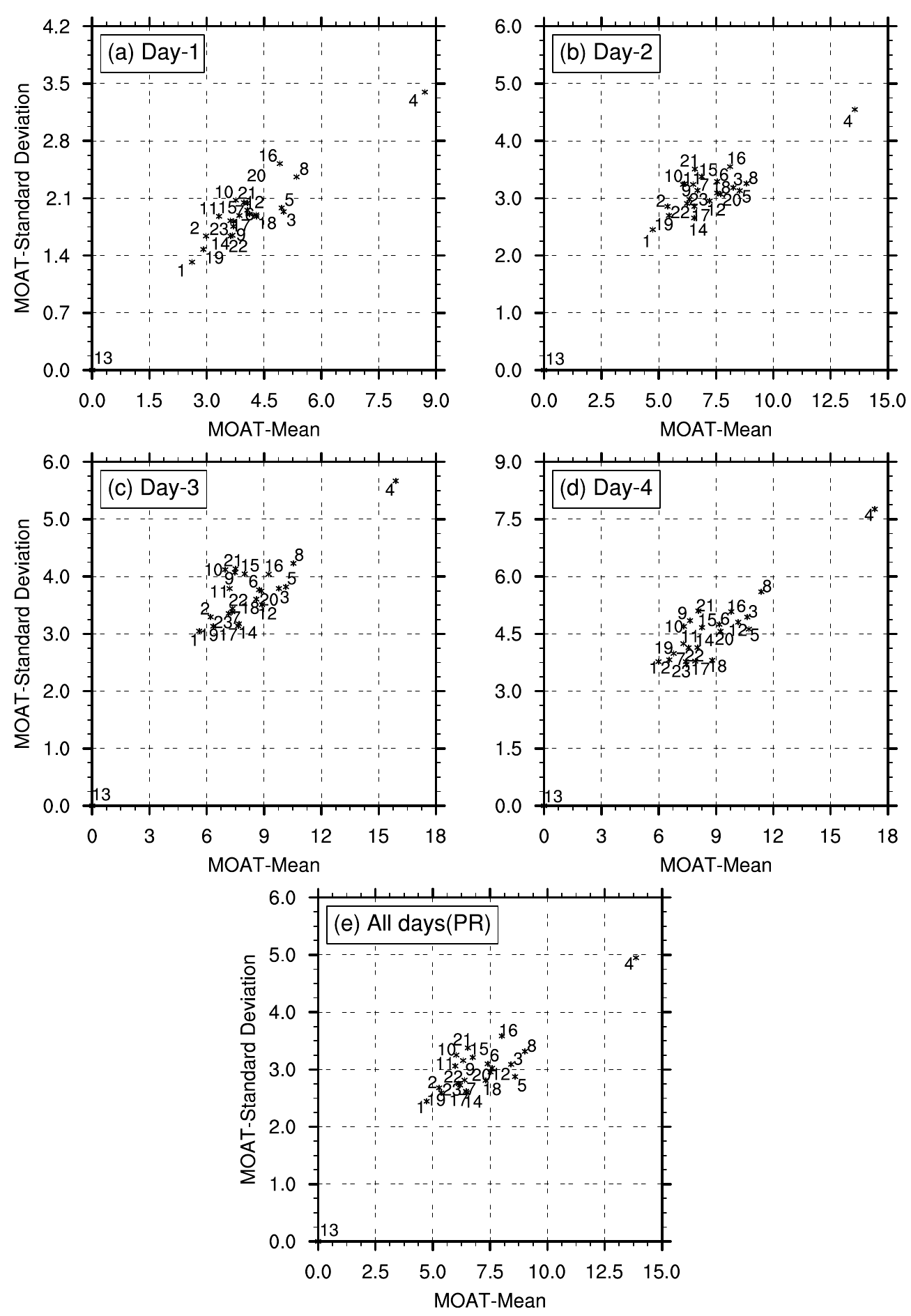}
    \caption{MOAT parameter sensitivity plots for precipitation corresponding to different lead times (a) to (d) lead times of one day to four days, respectively and (e) for all the days together}
    \label{fig:my_label3}
\end{figure}

\begin{figure}
\centering
    \includegraphics[scale=1]{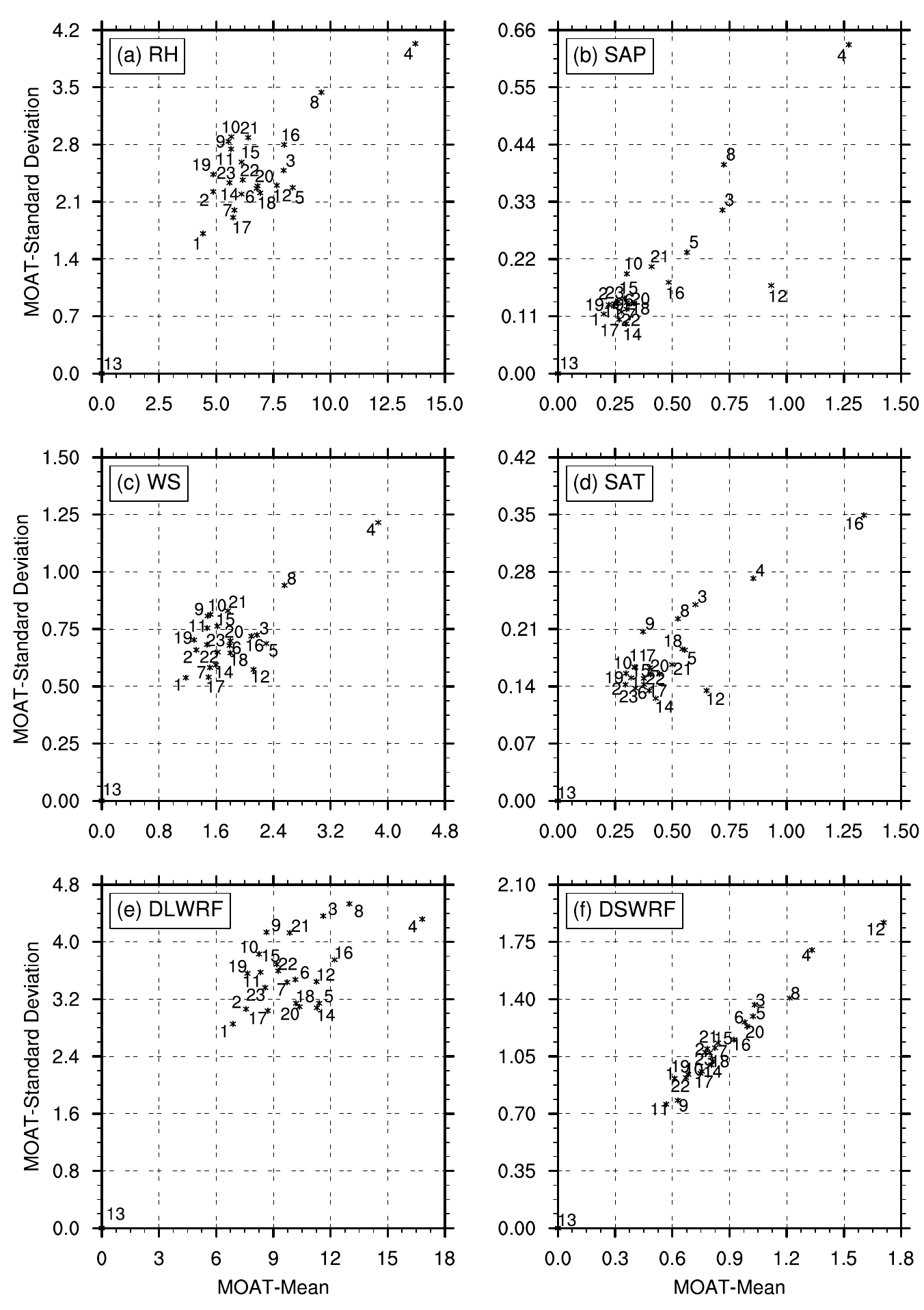}
    \caption{MOAT parameter sensitivity plots for surface meteorological variables (a) Relative Humidity, (b) Surface Air Pressure, (c) Wind Speed, (d) Surface Air Temperature, (e) Downward long-wave radiative flux, (f) Downward short-wave radiative flux}
    \label{fig:my_label4}
\end{figure}

\begin{figure}
\centering
    \includegraphics[scale=0.75]{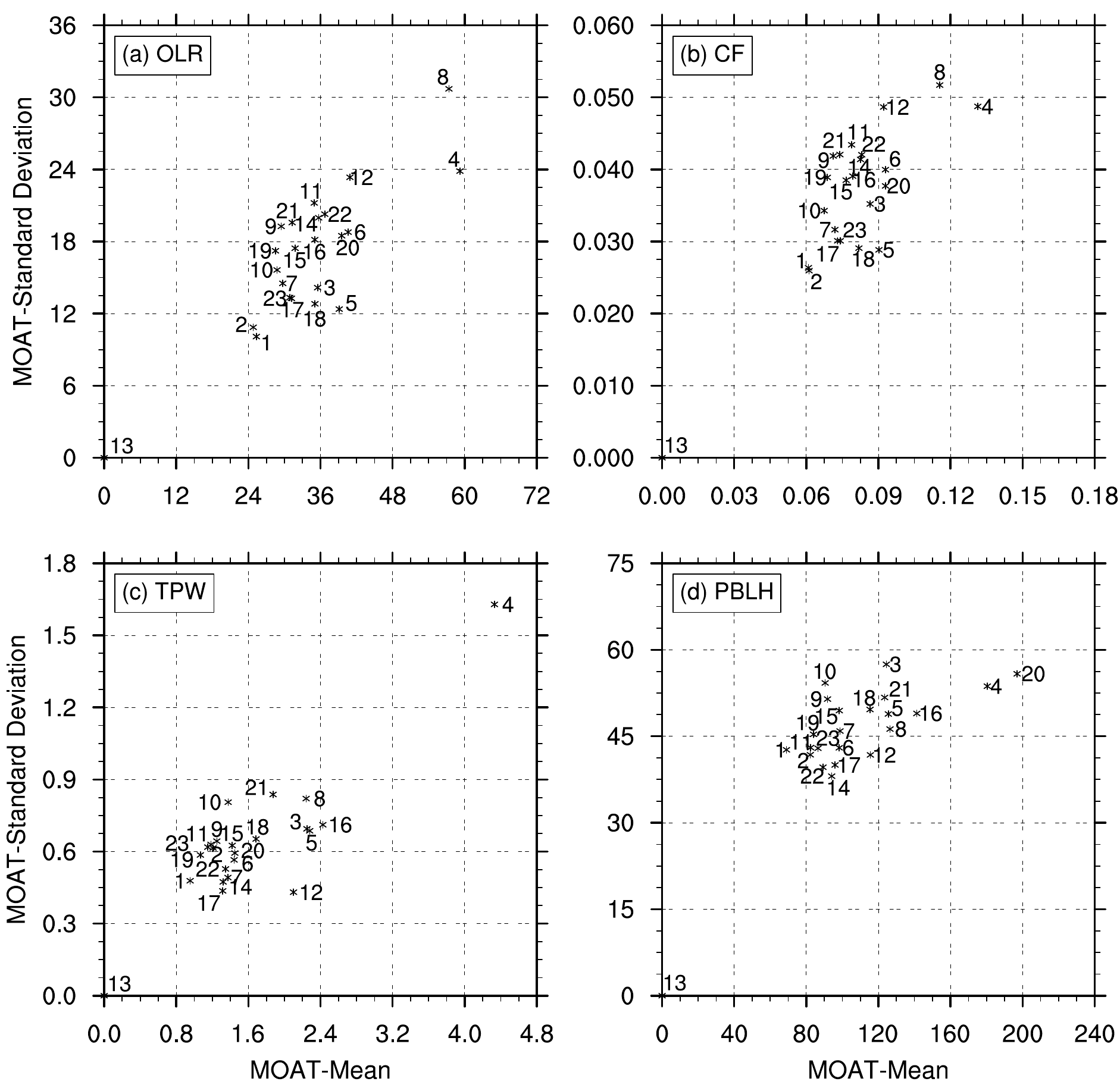}
    \caption{MOAT parameter sensitivity plots for atmospheric variables (a) Outgoing Long-wave Radiation, (b) Cloud Fraction, (c) Total Precipitable Water, (d) Planetary Boundary Layer Height}
    \label{fig:my_label5}
\end{figure}
MOAT sensitivity plots for the output variable precipitation are shown in Figure \ref{fig:my_label3}. Figures \ref{fig:my_label3}(a)-\ref{fig:my_label3}(d) correspond to the daily accumulated precipitation for day 1 to day 4 respectively, whereas Figure \ref{fig:my_label3}(e) corresponds to the average daily precipitation for all the days simulated. The x-axis and y-axis in each subplot correspond to the MOAT mean and MOAT standard deviation, respectively. It is observed that P4 has the highest MOAT mean value for all the days, which implies that this parameter influences precipitation the most, and P13 has the least MOAT mean and thus has no influence on the precipitation. Out of all the remaining parameters P8, P5, P3, and P16 have higher MOAT mean values compared to others which indicate that the precipitation is more sensitive to these parameters.\\

Similar sensitivity plots are generated for other surface meteorological variables represented in Figure \ref{fig:my_label4}. Figures \ref{fig:my_label4}(a)-\ref{fig:my_label4}(f) correspond to RH, SAP, WS, SAT, DLWRF, DSWRF respectively. P4 and P8 have higher MOAT mean values for RH, whereas P13 and P1 have lesser MOAT mean values. P4, P12, P3, and P8 have higher MOAT mean values for SAP, whereas P13 and P19 have lesser MOAT mean values. P4 and P8 have higher MOAT mean values for WS, whereas P13 and P1 have lesser MOAT mean values. P16 and P4 have higher MOAT mean values for SAT, whereas P13 and P2 have lesser MOAT mean values. P4, P8, and P16 have higher MOAT mean values for DLWRF, whereas P13 and P1 have lesser MOAT mean values. P12, P4, and P8 have higher MOAT mean values for DSWRF, whereas P13, P11, and P1 have lesser MOAT mean values. \\

Sensitivity plots are also obtained for atmospheric variables represented in Figure \ref{fig:my_label5}. Figures \ref{fig:my_label5}(a)-\ref{fig:my_label5}(d) correspond to OLR, CF,TPW and PBLH respectively. P4, P8, and P12 have higher MOAT mean values for OLR, whereas P13, P2, and P1 have lesser MOAT mean values. P4, P8, P20, P6, and P12 have higher MOAT mean values for CF, whereas P13, P2, and P1 have lesser MOAT mean values. P4, P16, P3, P5, P8, and P12 have higher MOAT mean values for TPW, whereas P13 and P1 have lesser MOAT mean values. P20, P4, and P16 have higher MOAT mean values for PBLH, whereas P13 and P1 have lesser MOAT mean values. \\

\begin{figure}
\centering
    \includegraphics[scale=0.5]{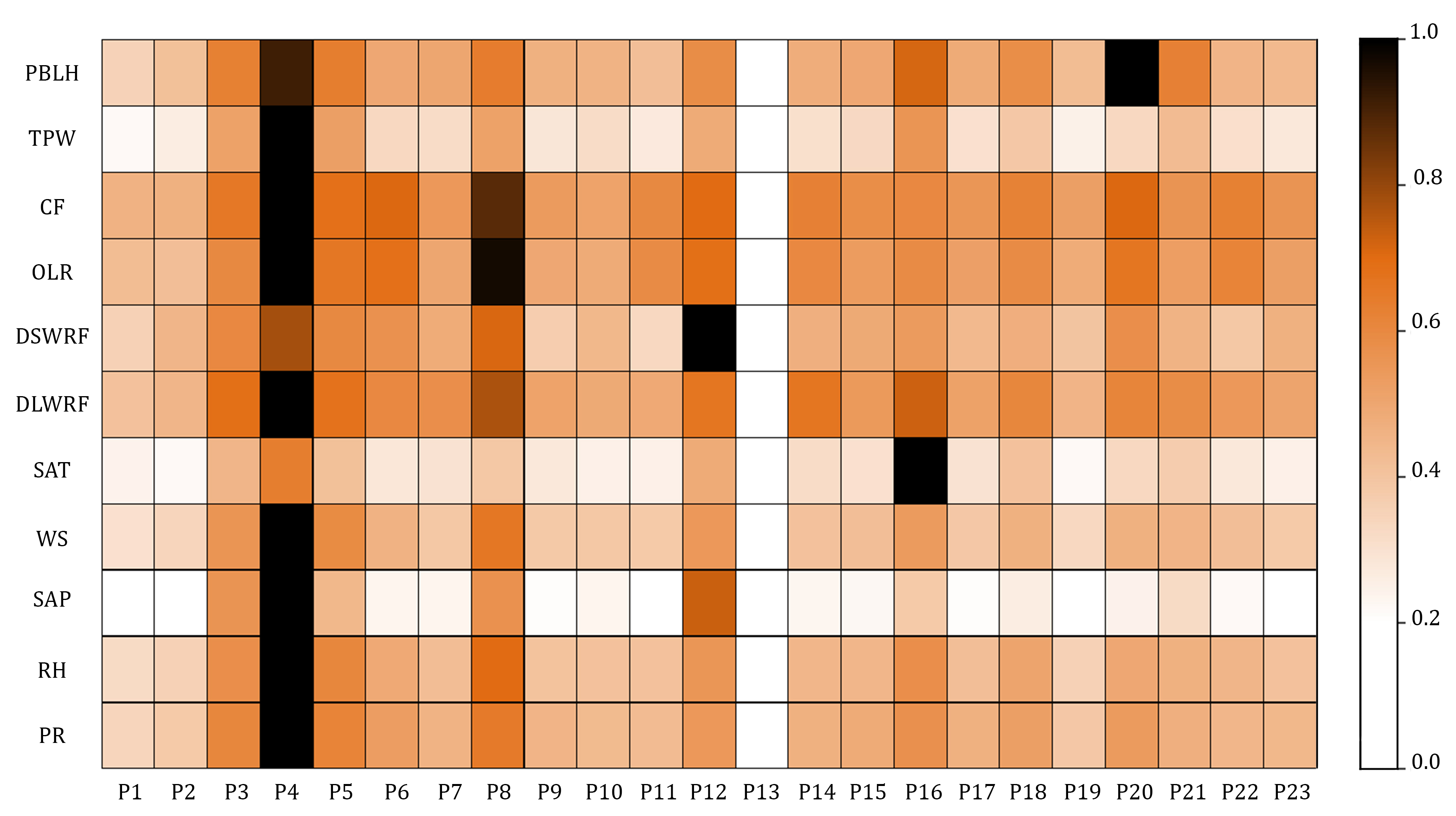}
    \caption{Heat map of normalized MOAT mean values for all the eleven variables evaluated against the 23 parameters}
    \label{fig:my_label6}
\end{figure}

MOAT mean values obtained for all the output variables considered for the 23 parameters are normalized and are plotted on a heat map, as shown in Figure \ref{fig:my_label6}. P4 is highly sensitive for almost all the output variables. Apart from P4, other parameters such as P3, P5, P8, P12, and P16 influence the output variables significantly. Four parameters namely P1, P2, P13, and P19 are observed to be least sensitive for all the output variables.\\
P4 is the multiplier for entrainment mass flux rate, which means that P4 influences the entrainment rate. If the entrainment rate increases, the moist convective core gets diluted, leading to a weak updraft mass flux resulting in lesser convective precipitation (PR). As P4 influences the strength of convection, this means that the formation of clouds also gets impacted. Therefore, atmospheric variables corresponding to clouds such as cloud fraction (CF), outgoing long-wave radiation at the top of the atmosphere (OLR), downward short-wave radiative flux (DSWRF) and downward long-wave radiative flux (DLWRF) are sensitive to P4. The intensity of convection also affects the total precipitable water (TPW). Precipitation results in the decrease of surface air temperature and an increase in relative humidity (RH) and a change in the surface air pressure (SAP). As a result of a change in the atmospheric circulation due to precipitation, wind speed (WS) also gets affected. P3 is the multiplier for downdraft mass flux rate which influences the downdraft. A higher value of P3 results in more downdraft leading to an increase in the evaporation of condensed water, resulting in lesser precipitation. P5 is the starting height of downdraft above the updraft source layer (USL). Downdraft flux starting at higher levels results in tall and narrow downdraft restricting the growth of convective precipitation. P3 and P5 also influence the strength of convection and thereby cloud formation similar to P4 which explains their impact on all the output variables.\\

The output variables corresponding to cloud formation such as cloud fraction (CF), outgoing long-wave radiation at the top of the atmosphere (OLR), downward short-wave radiative flux (DSWRF) and downward long-wave radiative flux (DLWRF) are sensitive to P8 which is the scaling factor applied to ice fall velocity. This controls the terminal velocity of the ice crystals that are descending, thereby influencing the cloud formation. As clouds(CF) reflect the solar radiation (DSWRF) and absorb and emit the long-wave radiation (DLWRF and OLR), this explains the sensitivity of cloud-related variables to P8. P12 is scattering tuning parameter which influences the scattering in the clear sky, thereby affecting the downward short-wave radiative flux (DSWRF) that reaches the ground. P16 is the multiplier for the saturated soil water content, which influences the transport of heat and moisture fluxes in the soil resulting in the heat exchange and evaporation between the land and atmosphere. Therefore, surface air temperature (SAT) is sensitive to P16. P20 is the critical Richardson number for the boundary layer of land which influences the planetary boundary layer height (PBLH).\\
\begin{figure}[H]
	\centering
	\includegraphics[scale=0.8]{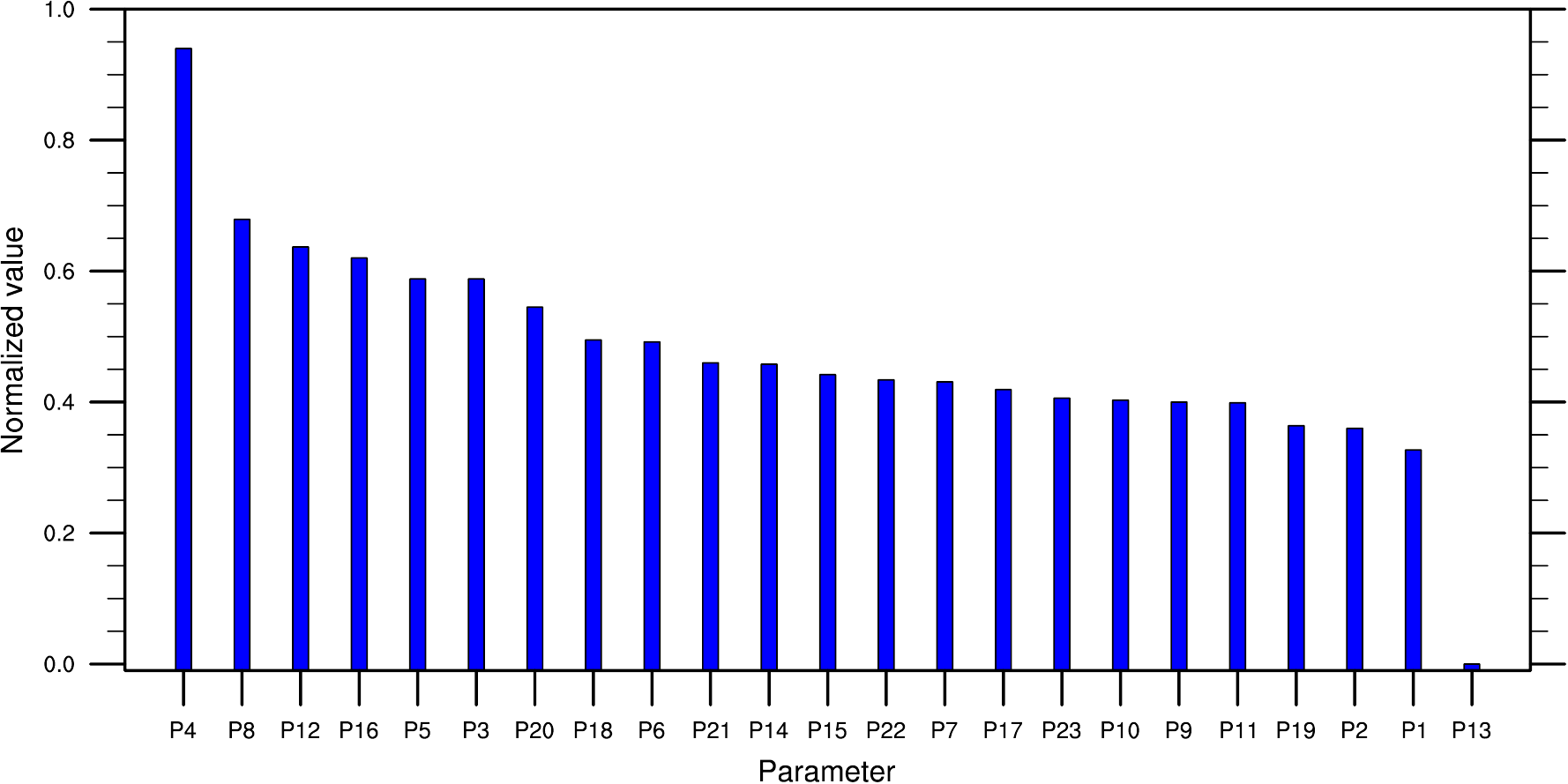}
	\caption{Comparison of average daily accumulated precipitation (mm) simulated over 48 days (Twelve 4-day events) using default and optimized parameters: (a) Observed, (b) Default, (c) Optimum, (d) Default - Observed, and (e) Optimum - Observed} 
	\label{fig:my_label6r}
\end{figure}

The parameters are ranked according to their senstivity and the ranks are presented in Figure \ref{fig:my_label6r}.\\

RMSE values for the output variables precipitation (PR), surface air temperature (SAT), surface air pressure (SAP) and wind speed at 10m (WS10) are evaluated for all the 240 parameter sets and also for the default parameter set. An improvement 15.14$\%$ (PR) and 15.58$\%$ (SAT) from the default parameter set is obtained. Minimal difference in RMSE is noted for the variables SAP and SAT of the optimum parameter set when compared to the default parameter set.\\

\begin{figure}
	\centering
	\includegraphics[scale=0.3]{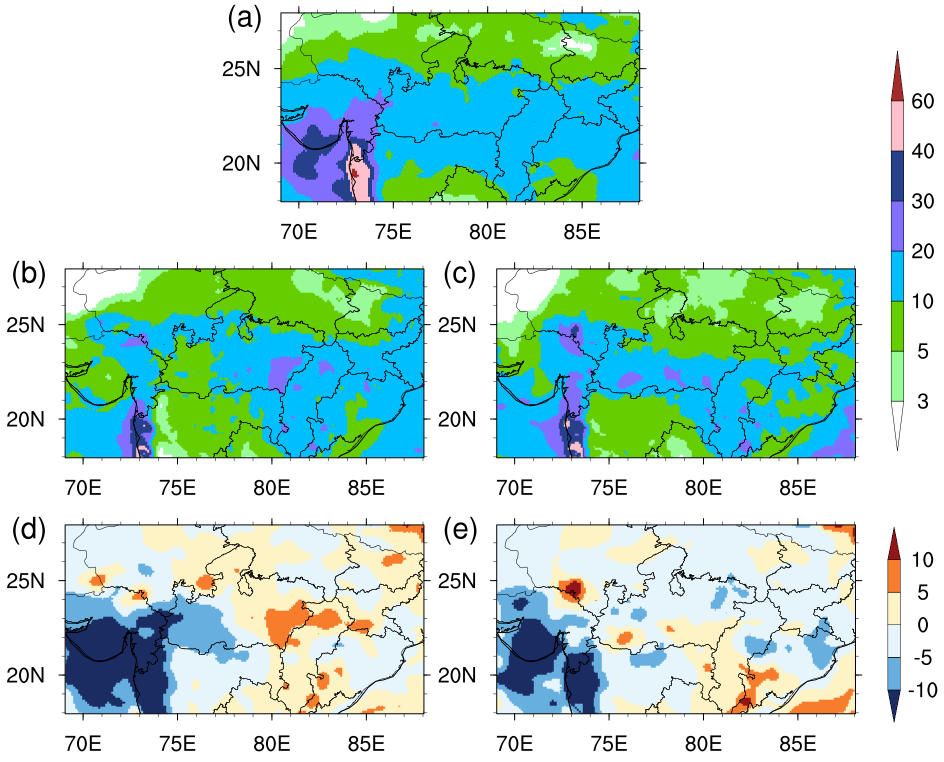}
	\caption{Comparison of average daily accumulated precipitation (mm) simulated over 48 days (Twelve 4-day events) using default and optimized parameters: (a) Observed, (b) Default, (c) Optimum, (d) Default - Observed, and (e) Optimum - Observed} 
	\label{fig:my_label7}
\end{figure}

\begin{figure}
	\centering
	\includegraphics[scale=0.3]{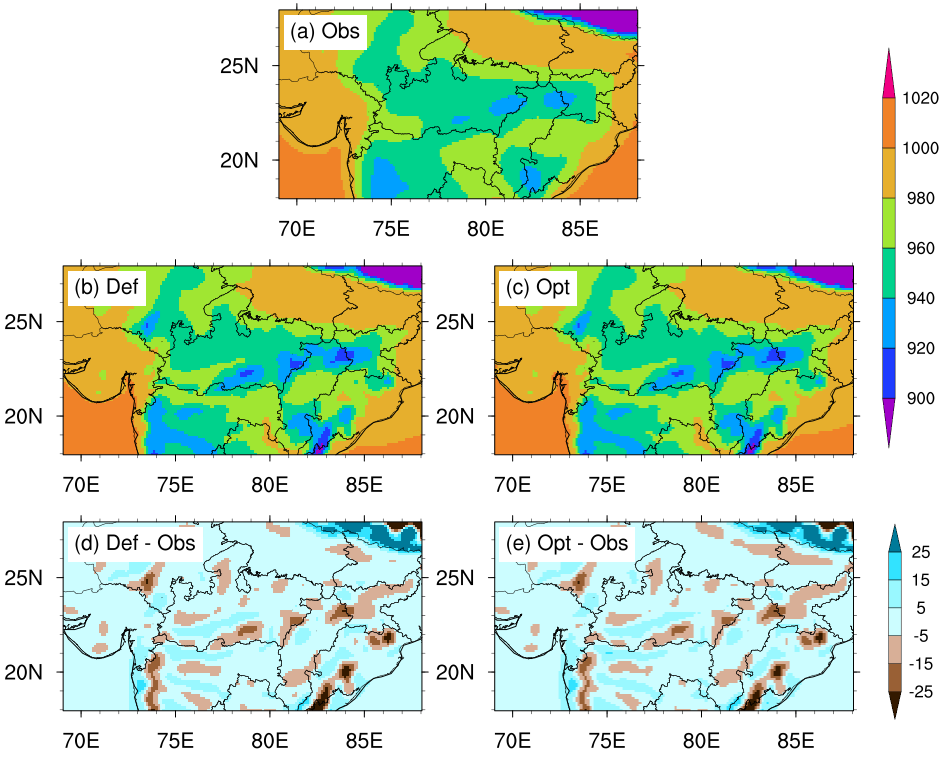}
	\caption{Comparison of average surface pressure (hPa) simulated over 48 days (Twelve 4-day events) using default and optimized parameters: (a) Observed, (b) Default, (c) Optimum, (d) Default - Observed, and (e) Optimum - Observed}
	\label{fig:my_label8}
\end{figure}

\begin{figure}
	\centering
	\includegraphics[scale=0.3]{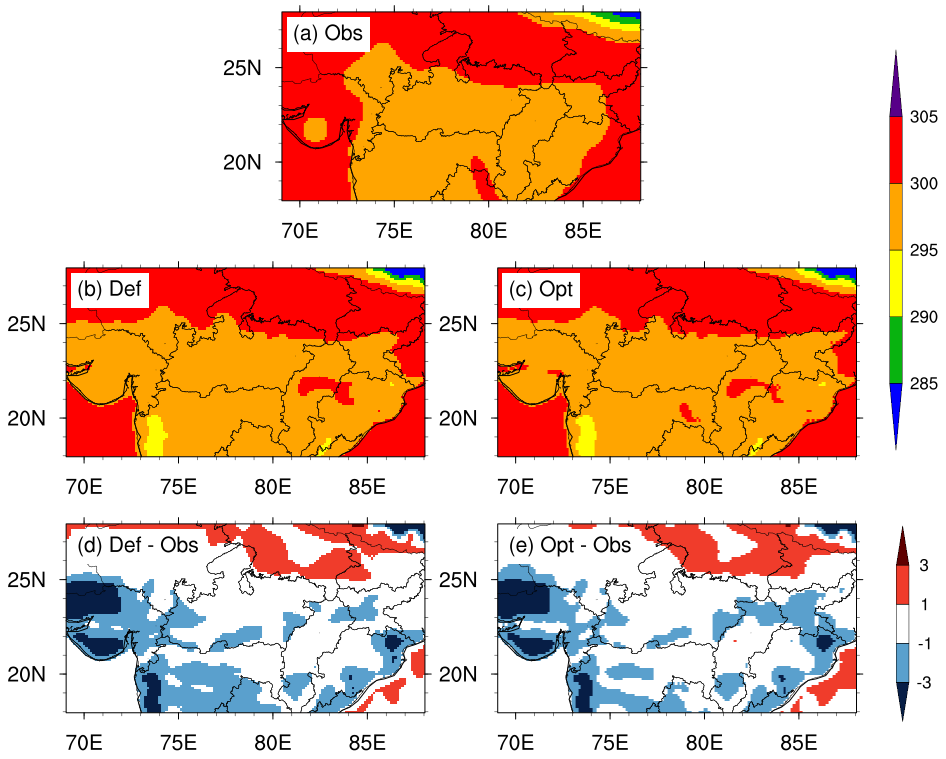}
	\caption{Comparison of average surface temperature (K) simulated over 48 days (Twelve 4-day events) using default and optimized parameters: (a) Observed, (b) Default, (c) Optimum, (d) Default - Observed, and (e) Optimum - Observed}
	\label{fig:my_label9}
\end{figure}

\begin{figure}
	\centering
	\includegraphics[scale=0.3]{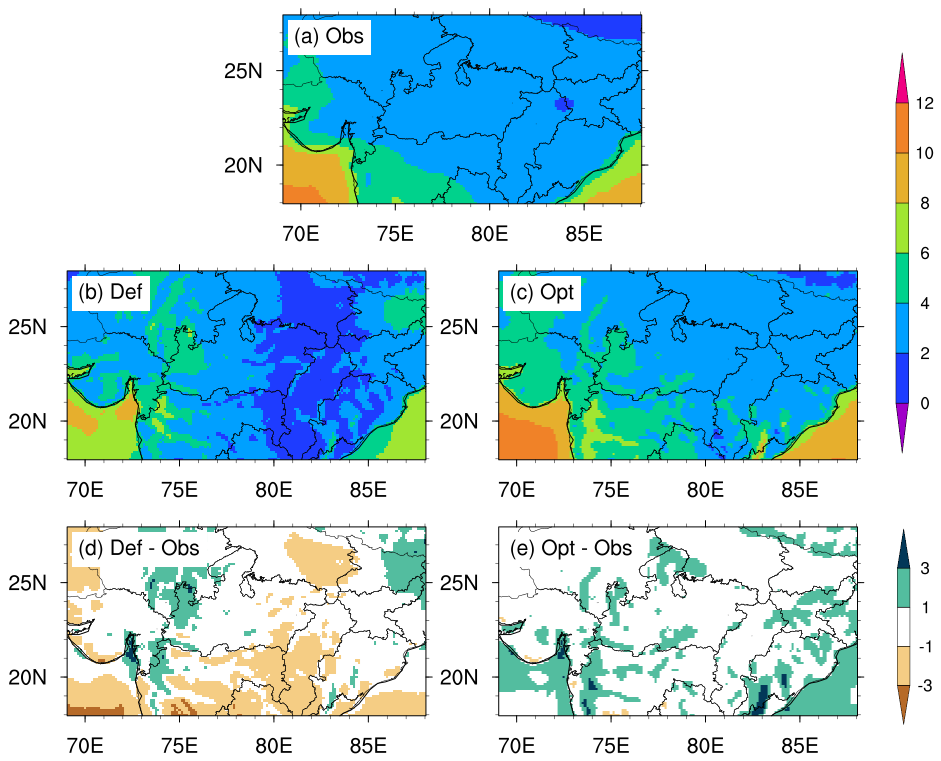}
	\caption{Comparison of average wind speed (m/s) at 10m simulated over 48 days (Twelve 4-day events) using default and optimized parameters: (a) Observed, (b) Default, (c) Optimum, (d) Default - Observed, and (e) Optimum - Observed}
	\label{fig:my_label10}
\end{figure}

The spatial patterns of the daily average of output variables PR, SAP, SAT and WS10 over 48 days (12 four-day events) for the optimum parameter (minimum RMSE out of 240) set and the default parameter set are compared against the observed data and is shown in Figures \ref{fig:my_label7}-\ref{fig:my_label10}. It can be seen that the output variables from the optimum parameter set are closer to the observed variables compared to the output variables from the default parameter set. It can be noted that the spatial pattern of PR and WS10 from the observed data is effectively simulated by the optimum parameter set which also justifies the high percentage of improvement of RMSE concerning the default parameter set.\\

\section{Conclusions}
\label{S:5}

The current study evaluated the sensitivity of eleven WRF model output variables and identified those parameters which have a higher influence on these outputs. Those parameters which have a very minimal effect on the model outcome are also vetted out in this process. The parametric study considered a total of 23 parameters corresponding to seven different physics schemes for the sensitivity analysis. Morris one at a time method, which is a global sensitivity analysis method, is utilized for the present parametric study. The sensitivity analysis is conducted over the monsoon core region during the Indian summer monsoon, which immensely affects the agricultural economy in India. The study was conducted for 12 high-intensity rainfall events during June, July, August, and September over the years 2015, 2016, and 2017.  \\

The study identified six parameters, namely P4, P8, P5, P12, P3, and P16, that have a significant impact on the output variables. Four of the 23 parameters, namely P13, P1, P2, and P19, are found to have a negligible effect on the model output. P4, in particular, has the highest sensitivity for all the variables, and P13 is found to have no impact at all. Furthermore, the RMSE values of the optimal solution obtained from the simulations conducted and the RMSE values from the default parameter set are compared for four observational variables. A significant improvement (around 15 \%) is observed for the output variables precipitation and wind speed. The parameters analyzed in this study are only a select few parameters in the WRF model. There may be other parameters to which the output variables may be more sensitive compared to the selected parameters.\\ 

Additionally, the sensitivity to the parameters in this study is obtained over the monsoon core region. The sensitivity of the output variables to the parameters may not remain the same for other areas of interest. The present parametric study can be considered as a precursor for further investigation of the sensitivities of other potential parameters for the Indian summer monsoon. Further, optimization of the sensitive parameters to accurately simulate the output variables can be done using advanced optimization techniques.

\vspace{2cm}
\bibliographystyle{model1}
\bibliography{sample.bib}

%\newpage
%\listoftables
%\listoffigures

\end{document}